%% file: main.tex
\documentclass[letterpaper,reprint]{article}

\usepackage[T1]{fontenc}

\usepackage{geometry}
\usepackage{setspace}

\usepackage[style = chem-acs,
articletitle=true]{biblatex}
\usepackage{graphicx}
\usepackage{float}
\newfloat{scheme}{htbp}{los}
\floatname{scheme}{Scheme}
\floatname{chart}{Chart}
\newfloat{graph}{htbp}{loh}

\usepackage{chemformula} 
\usepackage[version = 4]{mhchem} 

\usepackage{authblk}

\author[1]{Jānis Užulis*}
\author[1]{Andris Gulans}
\affil[1]{Department of Physics, Faculty of Science and Technology, University of Latvia, Jelgavas iela 3, Riga, LV-1004 Latvia}

\title{Hartree--Fock Limit for Energies in Solids}
\date{*Email: janis.uzulis@lu.lv}

\begin{document}

\maketitle

\begin{abstract}
This study establishes a route to the Hartree--Fock (HF) limit for molecules and solids within the linearized augmented plane wave (LAPW) framework.
We remove current limitations of the standard LAPW approach to nonlocal exchange by constructing radial basis functions and core orbitals consistently with the HF Hamiltonian.
The presented method yields total energies of molecules and solids with a precision of a few microhartrees, and we use it to provide reference data for 14 semiconductors and insulators.
For the systems considered in this study, the standard approach based on (semi)local potentials for constructing radial basis functions and core orbitals remains highly precise for practical relative energies, including molecular and solid-state formation energies and Si self-interstitial defect formation energies. 
More broadly, the results provide stringent all-electron benchmarks for basis-set and pseudopotential assessment, improve error control in hybrid-functional calculations within LAPW, and open the way to X-ray spectroscopy simulations within LAPW based directly on hybrid-functional core orbitals.
\end{abstract}


\section{INTRODUCTION}
Fock exchange is central to modern computational chemistry, entering both wave-function methods built on a Hartree--Fock (HF) reference and hybrid exchange-correlation functionals in density-functional theory (DFT).
The wave-function methods, such as the M{\o}ller--Plesset perturbation and coupled-cluster theories, offer a systematic route towards the exact many-electron solution of the Schr\"{o}dinger equation leading to reliable hard reference values where DFT approaches are insufficiently reliable.
They are
increasingly being used to describe solid-state and surface phenomena, including  
defect formation~\cite{Gruneis-2023-Si-defects,Simula-Si-defects-2026}, adsorption~\cite{Michaelides_Gruneis_SurfaceChem_2023}, 
chemical reactions on a surface~\cite{Weiske_and_Gruneis_Adsorption_reactions_2026},
and electronic excitations~\cite{Berkelbach_absorbtion_2021}.
Their predictive power depends not only on the underlying Hamiltonian but also on the quality of the one-electron basis.
While the basis-incompleteness issue is particularly severe in correlation-energy calculations, reliable results also require a well-converged HF reference.
Hybrid functionals often improve description of various properties compared to semilocal functionals, in particular, the band gaps of semiconductors and insulators, as the hybrids continue to improve and increase their predictive powers~\cite{Kronik_2021,Yang2023}.
As they bring predicted band gaps closer to experiment, numerical errors can become comparable to the remaining discrepancies, motivating further improvements in numerical precision~\cite{Uzulis2025}.

For molecular systems, the basis-incompleteness problem at the HF level has been largely resolved, as the absolute total energy can be determined with a predetermined number of significant digits.
Although absolute total energies are not usually the final target in applications, they provide the most stringent benchmark for basis completeness and implementation and thereby provide the input necessary to obtain reliable relative energies and derived properties.
There are finite-difference and finite-element discretizations for diatomic molecules  \cite{Jensen2005,Kobus2013,Lehtola2019}.
These methods allow total energies to be obtained with 1~$\mu$Ha precision selected as a numerical target well within any practical needs.
As a result, data from Ref.~\cite{Jensen2005} have been used as complete-basis reference for assessing Gaussian basis sets.
For larger molecules, multiresolution analysis (multiwavelet) approaches \cite{Harrison2004,MRChem2023} enabled
 all-electron HF calculations with a prescribed numerical precision.
Rather than relying on a hierarchy of predefined atom-centered basis sets, all these approaches use systematically refinable real-space representations.

In contrast, no similarly reliable route is available for solids, and benchmark-quality all-electron HF data are still limited. 
The multiwavelet approaches mentioned above have not been fully extended to periodic calculations.
Atom-centered basis sets based on linear combinations of atomic orbitals (LCAO), such as Gaussian-type orbitals (GTO), cannot be improved as systematically in periodic calculations as they can be for molecules because of quasi-linear dependencies and strongly system-dependent convergence behavior \cite{Maschio2024}. 
Efforts to optimize GTO basis sets for solids were clearly in need for a reference, and using plane-wave pseudopotential calculations for this purpose is not unusual~\cite{daga-2020,Maschio2024}.
While basis completeness is not a challenge in the latter type of calculations, all-electron total energies are typically not accessible and, furthermore, pseudization itself introduces additional errors.

The difficulty of obtaining HF-limit energies for solids is illustrated by the so-called LiH \textit{challenge}, originally posed by Gillan \textit{et al.} in 2008 as a benchmark for periodic HF methods and recently discussed in Ref.~\cite{Dovesi2026}.
Even for this system, whose primitive cell contains only two light atoms, approaching the HF limit requires numerically challenging large Gaussian basis sets, with some studies also relying on extrapolation procedures.
Even now, all-electron HF data for solids targeting  the complete-basis limit remain
extremely scarce and are essentially restricted to LiH \cite{paier-2009,usvyat-2011,daga-2020,sun-2023}.

In principle, all-electron linearized augmented plane waves (LAPW)~\cite{Slater1937,Andersen1975,SJOSTEDT200015}, the \textit{de facto} reference framework for solids~\cite{Lejaeghere2016}, should provide a controlled route to this limit. It was demonstrated for local density-functional theory, as Ref.~\cite{Gulans2018} showed near-$\mu$Ha agreement between LAPW and multiresolution analysis in molecular total energies. Recent developments, including the implementation of adaptively compressed exchange in LAPW, have already made periodic HF calculations substantially more practical by removing the explicit dependence of the Fock exchange evaluation on empty bands \cite{Zavickis_ACE, Uzulis2025}. However, reaching the HF complete-basis limit within this framework has still been prevented by a fundamental inconsistency: in standard LAPW implementations, radial basis functions and core orbitals are generated from a radial Schr\"odinger equation with a (semi)local potential, whereas the HF Hamiltonian contains nonlocal exchange.
Although error cancellation can make the standard PBE-based construction sufficient for meeting practical precision targets,
reliably quantifying the
remaining errors requires a method that can 
converge all-electron HF total energies systematically to within a few
$\mu$Ha of the complete-basis limit.
They can serve directly as benchmarks for validating other all-electron implementations, as it happened in the past in the context of other features.
For example, scalar-relativistic total energies computed with \texttt{exciting} were used in Ref.~\cite{ZORA_2024} for debugging and validating relativistic corrections implemented in a multiwavelet code.
Also, sufficiently converged total energies obtained with
different codes can be combined to compute energy
differences without relying on cancellation of numerical
errors, provided that the calculations use consistent
physical approximations. 

Basis incompleteness introduces errors also in orbital energies used for obtaining band gaps in hybrid-functional calculations.
A recent study used LAPW for providing benchmark gap values for solids and assessing quality of such data in literature~\cite{Uzulis2025}.
That study, however, relied on the standard approach with the PBE-derived radial basis and core orbitals and had to employ indirect ways for validation of the results.

In this paper, we remove this limitation by implementing HF-consistent radial basis functions and core orbitals in the all-electron LAPW code \texttt{exciting}~\cite{Gulans2014}. We validate the approach by computing absolute total energies within a few $\mu$Ha from the HF complete-basis limit for five molecules and comparing them with precise reference data. We then apply the method to crystalline solids, reassess existing LiH benchmark data, and provide high-quality HF reference energies and equation-of-state (EOS) parameters for LiH and 13 additional materials. To illustrate the broader utility of the method, we also report self-interstitial formation energies from HF calculations in bulk Si.
We use these reference calculations to examine error cancellation in energy differences obtained with the standard LAPW approach based on PBE-derived radial basis functions and core orbitals. 
Using NaCl as an example, we extend this analysis to the PBE0 band gap and the transition energy from the Na $2p$ core level to the conduction-band minimum.

\section{METHOD} 
Our method relies on the all-electron LAPW+lo framework that combines LAPWs with local orbitals~\cite{SJOSTEDT200015}. We briefly summarize its essential features and refer to Refs.~\cite{Singh2006,Gulans2014} for a more detailed description of the LAPW method.
The unit cell is divided into two types of regions: non-overlapping atomic spheres, or muffin-tins, and the interstitial region. In the latter, the valence wave functions are smooth and are expanded in a plane-wave basis as
\begin{equation}
\label{eq:lapwi}
    \psi_{n\mathbf{k}}(\mathbf{r})=\sum\limits_\mathbf{G} Z_{n\mathbf{G}+\mathbf{k}} e^{i(\mathbf{G}+\mathbf{k})\mathbf{r}},
\end{equation}
where $Z_{n\mathbf{G}+\mathbf{k}}$ are variational coefficients. Inside the muffin-tin centered at nucleus $\alpha$ located at $\mathbf{R}_\alpha$, the same wave functions are expressed as
\begin{equation}
\label{eq:lapwmt}
    \psi_{n\mathbf{k}}(\mathbf{r})=\sum\limits_{s\ell m} C_{n\mathbf{k},s\ell m} u_{s\ell}(r_\alpha) Y_{\ell m}(\hat{r}_\alpha),
\end{equation}
where $\mathbf{r}_\alpha=\mathbf{r}-\mathbf{R}_\alpha$,
and $Y_{\ell m}(\hat{r}_\alpha)$ denotes a spherical harmonic with $\ell$ and $m$
being the angular and magnetic quantum numbers, respectively.
Eq.~\ref{eq:lapwmt} permits several radial functions $u_{s\ell}(r)$ for each $\ell$ covered by the range of the index $s$.
$u_{s\ell}(r)$ correspond to atomic-like orbitals and are key to a compact yet accurate representation of the wave functions inside the muffin-tins. In practice, the radial basis functions and core orbitals are represented numerically on one-dimensional radial grids within each muffin-tin.

Within the nonrelativistic framework, the radial basis functions are chosen as the solutions $u_{\ell}(r;E_{\ell})$ of the radial Schr\"odinger equation
\begin{equation}
\label{eq:radial2}
    \left(-\frac{1}{2}\frac{\partial^2}{\partial r^2} + \frac{\ell(\ell+1)}{2r^2} + v_0(r) - E_{\ell}\right) r u_{\ell}(r;E_{\ell}) = 0,
\end{equation}
and their energy derivatives evaluated at $E=E_{\ell}$. Here, $v_0(r)$ denotes the spherically symmetric ($Y_{00}$) component of the local trial potential and, after convergence, of the self-consistent potential $v(\mathbf{r})$.
In practice, Eq.~\ref{eq:radial2} is transformed in two coupled first-order differential equations~\cite{Koelling1977} that are solved 
by integration using the initial condition based on the asymptotics at small $r$.
The potential $v_0(r)$ and the radial functions are represented on a grid, and the outward integration is carried out using the Numerov's or Runge-Kutta methods~\cite{Numerov1924,numerical-rec-book}, for example.
These methods provide a way of evaluating the function and its derivative at the next point if they are known at the current point.

Different radial functions $u_{s\ell}(r)$ for the same $\ell$ include solutions of Eq.~\ref{eq:radial2} at different energy parameters $E_\ell$ and their energy derivatives. For example, $u_\ell(r;E_\ell)$, $\dot{u}_\ell=\partial u_\ell/\partial E$, and $\ddot{u}_\ell=\partial^2u_\ell/\partial E^2$ carry different $s$ labels, as do functions evaluated at the semicore and valence energy parameters $E^\mathrm{sc}_{\ell}$ and $E^\mathrm{val}_{\ell}$, respectively.

The rapid and systematic convergence of total energies with respect to the radial basis follows from linearization of the radial solutions around selected energies $E_{\ell}$.
If the potential inside a muffin-tin is approximately spherically symmetric, each angular-momentum component of the wave function can be represented accurately using $u_{\ell}(r;E_{\ell})$ and a small number of energy derivatives, so that only a few radial functions per chemically relevant $\ell$ channel are typically required.
Using only the spherical symmetric component of the potential to generate the radial functions is a basis-construction approximation. The LAPW calculation that employs this basis relies on the full potential for generating Hamiltonian matrix elements and obtaining wave functions.
If $v(\mathbf{r})$ contains a substantial contributions beyond $v_0(r)$, it may be necessary to introduce additional radial functions in different $\ell$ channels or to choose smaller muffin-tin radii while maintaining a converged basis in the interstitial region.
Core orbitals, confined to the muffin-tin, are treated separately as products of a single radial function and a spherical harmonic.

In existing LAPW implementations, both the radial basis functions and the core orbitals are generated from the radial Schr\"odinger equation, Eq.~\ref{eq:radial2}, which assumes a local potential.
At the HF level, the nonlocal Fock exchange operator $\hat{K}$ transforms the problem from an ordinary differential equation to the an integro-differential equation and the aforementioned methods for outward integration are not directly applicable.
The usual approach is therefore to construct the radial functions from a fixed local potential, for example obtained from a preliminary PBE calculation \cite{Perdew1996,Betzinger_PBE0,Tran2011}.
For the light He and Be atoms, $\mu$Ha precision in the total PBE0 energy was demonstrated by systematically increasing the number of PBE-derived radial basis functions \cite{Zavickis_ACE}. 
The convergence of the total energy with respect to the number of radial functions is slow even for these atoms, and such an approach becomes impractical for heavier ones.
The number of radial functions required for this purpose, for example, for the oxygen atom, is prohibitively large, causing numerical instabilities.
Moreover, enlarging the radial basis does not address the inconsistency of core orbitals constructed from a local potential.

To remove this limitation, we construct the radial basis for valence electrons directly from the HF Hamiltonian. Employing spherical averaging within each muffin-tin, we express the action of the exchange operator on a basis function as
\begin{equation}
\label{eq:exchangemt1}
\hat{K}\left[u_{\ell}(r;E_{\ell})Y_{\ell m}(\hat{r})\right]=h_{\ell}(r)Y_{\ell m}(\hat{r}).
\end{equation}
The purpose of spherical averaging is to ensure that acting with $\hat{K}$ on a single spherical harmonic yields solely the same spherical harmonic.
The radial Schr\"odinger equation, Eq.~\ref{eq:radial2}, is then generalized to
\begin{equation}
\label{eq:radial_nonlocal}
\left(-\frac{1}{2}\frac{\partial^2}{\partial r^2} + \frac{\ell(\ell+1)}{2r^2} + v_0(r)-E_{\ell}\right) r u_{\ell}(r;E_{\ell}) = -r h_{\ell}(r),
\end{equation}
which we solve iteratively: (i) fix $h_{\ell}(r)$ and integrate Eq.~\ref{eq:radial_nonlocal} outwards for $u_{\ell}(r;E_{\ell})$, (ii) update $h_{\ell}(r)$ using Eqs.~\ref{eq:Fock_core} and \ref{eq:Fock_val}, and (iii) repeat steps (i) and (ii) until $u_{\ell}(r;E_{\ell})$ changes negligibly between successive iterations.

The radial basis functions are generated at predefined
energies,
whereas, for core orbitals, Eq.~\ref{eq:radial_nonlocal} must instead be solved as an eigenvalue problem with a decaying bound-state boundary condition.
We therefore obtain the core orbitals using the atomic solver of  Ref.~\cite{Uzulis2022}, which employs a Green's-function integral formulation similar to that used in multiresolution analysis~\cite{Harrison2004}.
Its adaptation to LAPW uses the local muffin-tin potential and exchange contributions from the occupied core and valence states of the LAPW calculation.
The valence contribution is evaluated using the muffin-tin density matrix as described below.
In this formulation, the core orbitals are obtained by inverting the differential operator $[\nabla^2+2\epsilon_{n\ell m}^{\mathrm c}]$ with $\epsilon_{n\ell m}^{\mathrm{c}}$ denoting the corresponding eigenvalue, or simply $[\nabla^2-\lambda^2]$ where $\lambda^2=-2\epsilon_{n\ell m}^{\mathrm{c}}$. 
Using the Green's function corresponding to the screened Poisson equation, the Schr\"{o}dinger equation can be rewritten as
\begin{equation}
\label{eq:int}
\psi_{n\ell m}^{\mathrm{c}}(\mathbf{r})=
-2 \int \frac{e^{-\lambda|\mathbf{r}-\mathbf{r}^\prime|}}{4\pi|\mathbf{r}-\mathbf{r}^\prime|} \left[v_0(r^\prime)+\hat{K}\right] \psi_{n\ell m}^{\mathrm{c}}(\mathbf{r}^\prime)d\mathbf{r}^\prime.
\end{equation}
The employed Green's function $e^{-\lambda|\mathbf{r}-\mathbf{r}^\prime|}/{4\pi|\mathbf{r}-\mathbf{r}^\prime|}$ corresponds to the decaying solutions and thus enforces the bound-state boundary condition.
Iterating Eq.~\ref{eq:int}, orthogonalizing the resulting wave functions, updating $\epsilon_{n\ell m}^{\mathrm{c}}$ and $\hat{K}$ yields the full set of core orbitals.

We point out that the outward integration approach described above for radial basis functions for valence electrons is not sufficiently robust for deep core orbitals.
To apply it for the eigenvalue problem, the outward integration has to be coupled with the shooting method~\cite{numerical-rec-book} which relies on counting the number of nodes for determining whether the orbital energy is below or above the eigenvalue.
However, the right-hand-side of Eq.~\ref{eq:radial_nonlocal} may artificially introduce additional nodes in $\psi_{n\ell m}^\textrm{c}(r)$ in the course of the iterative procedure and renders this approach unreliable. 
Finally, the core-electron problem can also be solved using a high-order finite-element method that yields atomic HF and hybrid-functional solutions with high numerical precision~\cite{Lehtola2019_atoms,CHENG2026}. 
Such a solver could be potentially integrated into LAPW code.

As the next step, we discuss how to evaluate the Fock exchange contribution entering Eqs.~\ref{eq:radial_nonlocal} and \ref{eq:int}.
Without loss of generality, we consider a muffin-tin centered at the origin and write
\begin{equation}
\label{eq:exchangemt2}
\hat{K} \left[u_{s\ell}(r)Y_{\ell m}(\hat{r})\right] =
-\frac{1}{N_\mathbf{k}}\sum_{n\mathbf{k}}\psi_{n\mathbf{k}}(\mathbf{r})
\int_{MT} \frac{\psi^\ast_{n\mathbf{k}}(\mathbf{r}^\prime) u_{s\ell}(r^\prime)Y_{\ell m}(\hat{r}^\prime)}{|\mathbf{r}-\mathbf{r}^\prime|}\,d\mathbf{r}^\prime.
\end{equation}
The summation index $n$ in Eq.~\ref{eq:exchangemt2} runs over both core and valence orbitals. Accordingly, we decompose the exchange contribution as $h_{s\ell}(r)=h^\mathrm{c}_{s\ell}(r)+h^\mathrm{val}_{s\ell}(r)$. The core contribution arises from the core orbitals with radial parts $\psi^{\mathrm{c}}_{n\ell}(r)$, which are fully localized within muffin-tins. 
It is given by
\begin{eqnarray}
\label{eq:Fock_core}
h^\mathrm{c}_{s\ell}(r) =
-\frac{1}{2}\sum_{n^\prime \ell^\prime }
(2\ell'+1)
\psi^{\mathrm{c}}_{n'\ell'}(r)
\sum\limits_{\ell''=|\ell-\ell'|}^{\ell+\ell'}{\!\!\!\!\!\!\!}{^\prime}
W_{\ell \ell' \ell''}
\int_0^{R_{\mathrm{MT}}} dr' \frac{r_{<}^{\ell''}}{r_{>}^{\ell''+1}} r'^2 \psi^{\mathrm{c}}_{n'\ell'}(r')u_{s\ell}(r'),
\end{eqnarray}
where
\begin{equation}
W_{\ell \ell' \ell''}=
\begin{pmatrix}
\ell & \ell' & \ell''\\
0 & 0 & 0
\end{pmatrix}^2
\end{equation}
is the Wigner 3-$j$ symbol squared, $r_< = \min(r,r^\prime)$, $r_> = \max(r,r^\prime)$, and $\sum^\prime$ denotes summation in steps of 2.

As evident from Eq.~\ref{eq:lapwmt}, the numerical representation of valence orbitals is compact.
Therefore, it is worthwhile to express the valence contribution in Eq.~\ref{eq:exchangemt2} using the density matrix defined as
\begin{equation}
\label{eq:density_matrix}
P_{s\ell m}^{s'\ell' m'}
=
\frac{1}{N_{\mathbf{k}}}
\sum_{n\mathbf{k}}\
C^\ast_{n\mathbf{k},s'\ell'm'}
C_{n\mathbf{k},s\ell m}.
\end{equation}
Eq.~\ref{eq:radial2} written for calculations with the local potential keeps only the spherically symmetric part of $v(\mathbf{r})$ to obtain the radial basis functions $u_{s\ell}(r)$ that do not depend on $m$.
To achieve the same result in the non-local case as well, we replace the density matrix by its spherical average:
\begin{equation}
P_{s\ell m}^{s'\ell' m'}\approx\overline{P}^{(\ell)}_{s s'}\delta_{\ell\ell^\prime}\delta_{mm^\prime}
\end{equation}
defined as
\begin{equation}
\label{eq:density_matrix_spherical}
\overline{P}^{(\ell)}_{s s'}
=
\frac{1}{2\ell+1}
\sum_{m=-\ell}^{\ell}
P_{s\ell m}^{s'\ell m} .
\end{equation}
This replacement ensures that the right-hand side of Eq.~\ref{eq:exchangemt1} can be expressed using a single spherical harmonic.
The valence contribution to the action of the exchange operator on the target function $u_{s\ell}(r)Y_{\ell m}(\hat{\mathbf r})$ is
\begin{equation}
\label{eq:Fock_val}
\begin{aligned}
h^\mathrm{val}_{s\ell}(r)
=-\frac{1}{2}
\sum_{s's''\ell'}
(2\ell'+1)\,
\overline{P}^{(\ell')}_{s's''}\,
u_{s'\ell'}(r)
\sum_{\ell''=|\ell-\ell'|}^{\ell+\ell'}{\!\!\!\!\!\!\!}{^\prime}
W_{\ell\ell'\ell''}
\int_0^{R_{\mathrm{MT}}} dr'\,
\frac{r_<^{\ell''}}{r_>^{\ell''+1}}\,
r'^2 u_{s''\ell'}(r')u_{s\ell}(r').
\end{aligned}
\end{equation}

The radial basis is constructed with the core--valence partitioning of each atomic species taken into account. For an $\ell$ channel containing a single valence shell, we use at least three radial degrees of freedom: $u_{\ell}(r;E^\mathrm{val}_{\ell})$, $\dot{u}_{\ell}(r;E^\mathrm{val}_{\ell})$, and $\ddot{u}_{\ell}(r;E^\mathrm{val}_{\ell})$. If both semicore and valence shells are present in the same $\ell$ channel, we use at least five radial functions: $u_{\ell}(r;E^\mathrm{sc}_{\ell})$, $\dot{u}_{\ell}(r;E^\mathrm{sc}_{\ell})$, $\ddot{u}_{\ell}(r;E^\mathrm{sc}_{\ell})$, $u_{\ell}(r;E^\mathrm{val}_{\ell})$, and $\dot{u}_{\ell}(r;E^\mathrm{val}_{\ell})$. The energy parameters $E^\mathrm{sc}_{\ell}$ and $E^\mathrm{val}_{\ell}$ are chosen using the Wigner--Seitz rule~\cite{Wigner-Seitz_rule}, i.e., as averages of the energies for which $du_\ell/dr=0$ and $u_\ell=0$ at $r=R_\mathrm{MT}$. This set is typically sufficient in materials science applications. In benchmark calculations targeting the complete-basis limit, the described approach works well for sufficiently small muffin-tins. In practice, however, larger muffin-tins are often preferable to reduce the number of LAPWs, and, in such cases, we complement the radial basis by additional high-energy atomic-like orbitals~\cite{Michalicek_2013,Zavickis_ACE}.

\section{MOLECULES}
\subsection{Basis Completeness}
Since the LAPW+lo basis combines plane waves in the interstitial region with atomic-like functions inside the muffin-tins, both contributions must be converged. The quality of the numerical representation in the interstitial region is controlled by a dimensionless cutoff parameter $R_\mathrm{MT}K_\mathrm{max}$, where $R_\mathrm{MT}$ and $K_\mathrm{max}$ are the atomic-sphere radius and the largest allowed $|\mathbf{G}+\mathbf{k}|$, respectively.
We observe the usual exponential convergence of the HF total energy with respect to the cutoff (see Supporting Information), and this is consistent with previous LAPW studies \cite{Gulans2014,Gulans2018,Zavickis_ACE}.
Based on these convergence tests, we set $R_\mathrm{MT}K_\mathrm{max}=8$, $13$, and $13$ for H, F, and Cl, respectively.
For a heteronuclear molecule, different values of $R_\mathrm{MT}K_\mathrm{max}$ are obtained by using species-dependent muffin-tin radii while retaining a single global value of $K_\mathrm{max}$.

\begin{figure}[H]
  \centering%
  \includegraphics[width=240pt]{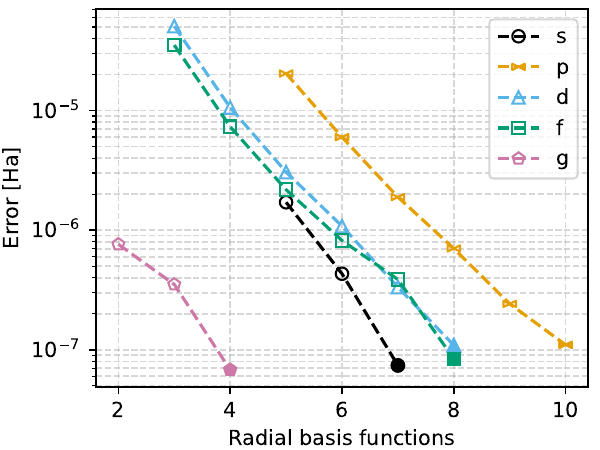}  
  \caption{\label{fig:lo_conv}Convergence of the total HF energy (in Ha) depending on the number of radial basis functions in each $\ell$ channel for the Cl$_2$ molecule with $R_\mathrm{MT}=1.6~a_\mathrm{B}$, where $a_\mathrm{B}$ is the Bohr radius.
  The filled markers indicate the radial-basis settings used in the calculation whose total energy is reported in Tab.~\ref{tab:molecules}.
  }
\end{figure}

Having observed the usual behavior with respect to the $R_\mathrm{MT}K_\mathrm{max}$ value, we focus on the convergence of the atomic-like basis within the muffin-tins.
Fig.~\ref{fig:lo_conv} shows the convergence of the total energy versus the atomic-like basis for Cl$_2$. Despite the ground-state Cl atom formally containing only $s$- and $p$-occupied shells, additional $d$ and $f$ radial functions are needed to reach $\mu$Ha precision in a molecular calculation.
In contrast to the results obtained with PBE-derived radial functions,
(see the lower panel of Fig.~2 in Ref.~\cite{Zavickis_ACE}), we observe a rapid reduction of the total-energy error with increasing radial-basis size, thereby confirming the efficiency of the present HF-consistent basis construction.
A direct comparison with PBE-derived radial basis functions for Cl$_2$ is complicated by the special treatment of the Cl $1s$ as a core orbital.
With a PBE-derived core, enlarging the valence basis cannot eliminate the core-orbital error and therefore cannot recover the HF complete-basis limit.
Combining a HF core with a PBE-derived radial basis, however, does not automatically ensure core--valence orthogonality, and this combination can yield nonvariational energies below the complete-basis limit.
For this reason, we carry out a comparison of PBE- and HF-derived radial bases for diamond with all electrons described on the valence or semicore level and refer to Fig.~S2 of the Supporting Information.

\subsection{Total Energies -- Validation \label{sec:validation}}

To validate the present HF-consistent LAPW basis, we compute nonrelativistic HF total energies for five diatomic molecules (H$_2$, FH, F$_2$, ClF, and Cl$_2$) and compare them with high-precision reference data from the literature~\cite{Jensen2005,Lehtola2019} or from additional calculations using the \textsc{HelFEM} code~\cite{Lehtola_HelFEM_git} consistently with Ref.~\cite{Lehtola2019}. The molecular geometries are taken from the corresponding reference studies, except for H$_2$, for which we use the bond length of 0.7414~\AA{} as in Ref.~\cite{Bischoff2009}. To eliminate interactions between periodic images, all calculations are performed for large cubic unit cells employing a spherical cutoff in both the Coulomb and exchange terms~\cite{Spencer2008,Tran2011}.
In addition to eliminating basis-incompleteness errors, our molecular calculations must be converged with respect to the unit-cell size. We ensure this by systematically increasing the dimensions of the unit cell with a step of $5~a_\mathrm{B}$ until the change in the total energy falls below the target numerical precision of 1~$\mu$Ha.
As shown in Tab.~\ref{tab:molecules}, the LAPW results are in excellent agreement with the reference values, with a maximum deviation of only 2.5~$\mu$Ha.

\begin{table}[H]
\caption{\label{tab:molecules}Total nonrelativistic HF energies of diatomic molecules computed with \texttt{exciting} and compared with reference values from the literature or from an additional \textsc{HelFEM} calculation.}
\centering
\begin{tabular}{llll}
\hline
\hline
Molecule & \multicolumn{1}{c}{\texttt{exciting}\textsuperscript{\emph{a}} [Ha]} & {\textsc{HelFEM}\textsuperscript{\emph{b}} [Ha]} & \multicolumn{1}{c}{$\Delta E~[\mathrm{\mu Ha}]$} \\
    \hline
H$_2$       & ~~~-1.133624 & ~~~-1.133624\textsuperscript{\emph{c}}   & -0.3        \\
FH       & -100.070801  & -100.070803 & ~1.4      \\
F$_2$    & -198.773442  & -198.773445 & ~2.3      \\
ClF      & -558.917625  & -558.917626 & ~0.9      \\
Cl$_2$   & -919.008932  & -919.008935 & ~2.5      \\
\hline
\hline
\end{tabular}

\textsuperscript{\emph{a}}Present work;\\
\textsuperscript{\emph{b}}Ref.~\cite{Lehtola2019} if not noted otherwise;\\
\textsuperscript{\emph{c}}Obtained in the present work using \textsc{HelFEM}.
\end{table}

\subsection{Relative Energies}

\begin{figure}[H]
    \centering
    \includegraphics[width=360pt]{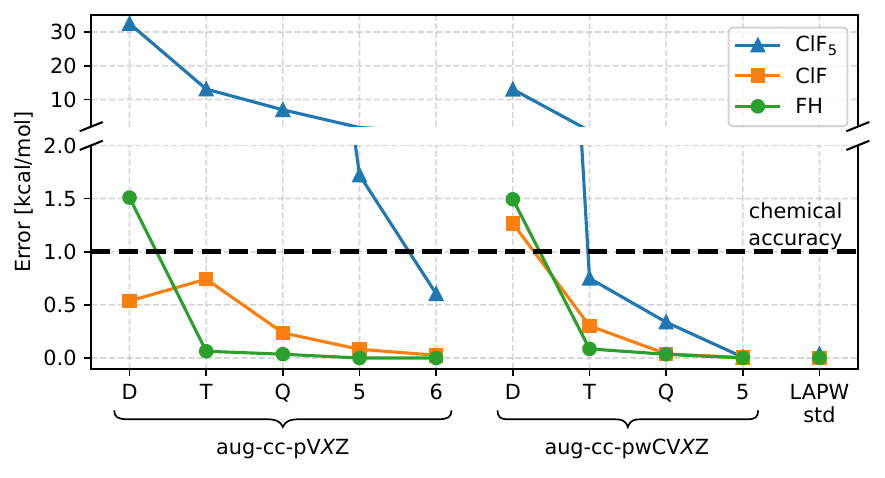}
    \caption{Absolute formation energy errors (in kcal/mol) in GTO calculations using the hierarchy of basis sets from aug-cc-pVDZ to aug-cc-pV6Z, 
    from aug-cc-pwCVDZ to aug-cc-pwCV5Z and in the 
    LAPW runs (labeled with \texttt{LAPW std}). The latter calculations employ PBE-derived core orbitals, PBE-derived radial basis, and a reduced cutoff $R_\mathrm{MT}K_\mathrm{max}$. The reference for the error is the LAPW calculation reported in Tab.~\ref{tab:molecules}. For ClF$_5$, whose total energy is not included in that table, the converged LAPW HF value used is $-956.490101$~Ha. The corresponding formation energies for all three molecules are reported in Tab.~S1 of the Supporting Information. 
    The vertical axis is broken to display errors occurring on substantially different energy scales within the same figure. 
    Lines connecting the data points are shown as a guide to the eye.}
       \label{fig:molecule-reaction-errors}
\end{figure}

Although the HF limit for absolute total energies requires a highly converged basis, relative-energy calculations benefit from error cancellation and can therefore often be obtained with high quality using a less precise basis.
With reference data obtained, we assess the performance of the LAPW(PBE/PBE) setup by calculating formation energies for FH, ClF, and ClF$_5$ and comparing them with the fully HF-consistent reference LAPW(HF/HF). Here and below, LAPW(A/B) denotes calculations using A-derived core orbitals and B-derived radial basis functions, where A and B denote either HF or PBE.
The reference LAPW(HF/HF) calculations employ more stringent $R_\mathrm{MT}K_\mathrm{max}$ values, whereas LAPW(PBE/PBE) is also tested with reduced cutoffs. The formation energies are defined as the HF energy differences between the considered molecule and the corresponding homonuclear diatomics. To put the resulting errors into context, we compare them with those from GTO calculations using the aug-cc-pV$X$Z basis sets~\cite{Dunning1989,Kendall1992,Woon1993} with $X=2$--6 
as well as aug-cc-pCV$X$Z and aug-cc-pwCV$X$Z~\cite{Woon1995,Peterson2002} with $X=2$--5.
The ClF$_5$ geometry was taken from the NIST CCCBDB database \cite{CCCBD_nist}.
As shown in Fig.~\ref{fig:molecule-reaction-errors}, the errors introduced by the LAPW(PBE/PBE) setup remain very small for all three systems, whereas the GTO results show substantially stronger basis-set dependence, particularly for ClF$_5$ 
using the aug-cc-pV$X$Z basis set.

For FH and ClF, formation-energy errors are below 1~kcal/mol at $X\geq 3$ level for all three Gaussian basis families considered.
ClF$_5$ is markedly more challenging for the aug-cc-pV$X$Z series, with even aug-cc-pV6Z differing by about 0.6~kcal/mol from the converged LAPW reference.
The core-valence basis families, aug-cc-pwCV$X$Z and aug-cc-pCV$X$Z, yield errors below 1~kcal/mol for ClF$_5$ starting from $X=3$ and 4, respectively.
At the 5Z level, both families agree with LAPW within 0.01~kcal/mol.
We explain this improved performance of the core-valence basis sets compared to aug-cc-pV$X$Z by the additional degrees of freedom allocated for the core that are needed to describe the high oxidation state in ClF$_5$.
The LAPW(PBE/PBE) setup yields errors below 0.04~kcal/mol for all three molecules even with the reduced cutoff.

\section{SOLIDS}
\subsection{HF reference energies of crystalline solids}
Having validated the HF-consistent LAPW basis for molecules, we now apply it to crystalline solids. We compute the HF limit of total and cohesive energies for 14 elemental and binary semiconductors and insulators in their experimental structures, spanning the first three rows of the periodic table. 

To limit numerical errors in solid-state HF calculations, one has to address basis incompleteness as in the molecular case and choose a sufficiently dense $\mathbf{k}$-point mesh used for Brillouin-zone sampling.
The plane-wave basis in the interstitial region results in a systematic exponential convergence of the total energy with respect to the LAPW cutoff.
To improve the quality of the basis within the muffin-tin region, we increase the number of radial functions following the same approach as shown in Fig.~\ref{fig:lo_conv}.
The basis-incompleteness errors in the molecular and solid-state cases behave in a similar way.
The major difference between them is in the additional complication in periodic HF calculations requiring to handle the singularity in 
the integral 
\begin{equation}
\hat{K} \psi_{n\mathbf{k}}(\mathbf{r}) =
-\frac{1}{N_\mathbf{k}}\sum_{n^\prime\mathbf{k}^\prime}\psi_{n^\prime\mathbf{k}^\prime}(\mathbf{r}) \int \frac{\psi^\ast_{n^\prime\mathbf{k}^\prime}(\mathbf{r}^\prime) \psi_{n\mathbf{k}}(\mathbf{r}^\prime)} {|\mathbf{r}-\mathbf{r}^\prime|}\,d\mathbf{r}^\prime,
\end{equation}
when $\mathbf{k}=\mathbf{k}^\prime$ and $n=n^\prime$.
We address it by using the spherical truncation of the Coulomb potential following the approach introduced by Spencer and Alavi in Ref.~\cite{Spencer2008} and systematically increasing the density of the $\mathbf{k}$-point mesh until the total energy reaches the target numerical precision, as documented in the Supporting Information.
Based on these convergence tests and the results from the molecular validation in Tab.~\ref{tab:molecules}, we estimate the numerical uncertainty in the solid-state total energies to be approximately 3~$\mu$Ha per unit cell.

\begin{table}[H]
\caption{\label{tab:solids} Structural parameters (in \AA{}) used in the calculations, HF total energies per unit cell (in Ha), cohesive energies per formula unit (in kcal/mol) and the transition energies (in eV) corresponding to the HF fundamental band gap, if not noted otherwise.
The structure of each material is given with the following acronyms: RS -- rocksalt, WUR -- wurtzite, ZB -- zinc blende, and DIA -- diamond.}
\centering
\begin{tabular}{llrrrc}
\hline
\hline
Material & $a$ & \multicolumn{1}{c}{$E_\mathrm{tot}$} & \multicolumn{1}{c}{$E_\mathrm{coh}$}  & \multicolumn{1}{c}{$E_\mathrm{g}$}  & \multicolumn{1}{c}{Transition}  \\
\hline
LiH (RS)  & 4.084  & -8.064710 & -82.806 & 10.897 & $\mathrm{X}\rightarrow\mathrm{X}$\\
LiF (RS)  & 4.01   & -107.094276 & -153.877& 22.022& $\Gamma\rightarrow\Gamma$\\
BeO (WUR) & 2.6983\textsuperscript{\emph{a}} & -179.477379 & -217.549 & 18.990& $\Gamma\rightarrow\Gamma$\\
BN (ZB)   & 3.616  & -79.284606 & -217.682& 13.723& $\Gamma\rightarrow\mathrm{X}$\\
BP (ZB)   & 4.5383 & -365.503892 & -157.792& 7.957& $\Gamma\rightarrow\mathrm{X}$\textsuperscript{\emph{c}}\\
C (DIA)   & 3.567  & -75.775368 & -121.702& 13.248& $\Gamma\rightarrow\mathrm{X}$\textsuperscript{\emph{c}}\\
NaF (RS)  & 4.62   & -261.487054 & -132.902& 18.477& $\Gamma\rightarrow\Gamma$\\
NaCl (RS) & 5.595  & -621.536214 & -117.537& 13.967& $\Gamma\rightarrow\Gamma$\\
MgO (RS)  & 4.207  & -274.701730 & -168.244& 15.718& $\Gamma\rightarrow\Gamma$\\
MgS (RS)  & 5.1913 & -597.333241 & -128.775& 10.392& $\Gamma\rightarrow\mathrm{X}$\\
AlN (WUR) & 3.11\textsuperscript{\emph{b}} & -593.148088 & -181.136& 13.535& $\Gamma\rightarrow\Gamma$\\
AlP (ZB)  & 5.451  & -582.801134 & -126.142& 7.823& $\Gamma\rightarrow\mathrm{X}$\\
SiC (ZB)  & 4.358  & -326.886747 & -209.632& 8.664& $\Gamma\rightarrow\mathrm{X}$\\
Si (DIA)  & 5.43   & -577.941656 & -70.212& 6.554& $\Gamma\rightarrow\mathrm{X}$\textsuperscript{\emph{c}}\\
\hline
\hline
\end{tabular}

\textsuperscript{\emph{a}} c=4.3776~\AA{} and z=0.3782; \\
\textsuperscript{\emph{b}} c=4.9791~\AA{} and z=0.3821;\\
\textsuperscript{\emph{c}} The $\Gamma\rightarrow\mathrm{X}$ transition energy is provided instead of the band gap as the conduction-band minimum is on the $\Gamma-\mathrm{X}$ line at a low-symmetry point. 
\end{table}

Tab.~\ref{tab:solids} reports the resulting HF total energies per unit cell and cohesive energies per formula unit.
Expressing the cohesive energy requires the total energies of open-shell atoms.
As our HF code currently supports only spin-restricted calculations, we use unrestricted-HF calculations in NWChem code~\cite{NWChem2010} to obtain the atomic reference energies.
Such an approach works only because we reach absolute convergence with both codes and do not rely on a cancellation of errors.
For the LCAO calculations, we construct an even-tempered Gaussian basis similarly to how it was done in Ref.~\cite{Gulans2018} and obtain total energies within 1--2~$\mu$Ha of the HF limit.
A comparison of the atomic energies for Li, Be, N, Na, Mg, and P between the LCAO calculations and our atomic solver~\cite{Uzulis2022} reveals a sub-$\mu$Ha level of agreement.

The present results therefore provide a broader set of reference-quality all-electron HF energies for periodic materials, as direct HF total-energy benchmarks for solids remain extremely scarce. 
Published direct all-electron total-energy benchmarks are essentially limited to LiH, while for diamond C we found only one published total-energy value presented as the complete-basis estimate~\cite{daga-2020}.
Tab.~\ref{tab:lih_tot.vs.literature} compares the total energies for LiH and C from the present work with previously published data.
Earlier direct total energies for LiH were obtained only with LCAO-based approaches and fall between $-8.064539$~Ha and $-8.064618$~Ha \cite{paier-2009,civalleri-2010-paier-comment,Dovesi2026,usvyat-2011,daga-2020,sun-2023}.
Extrapolated or perturbative HF-limit estimates extend to lower energies, and the perturbative estimate of Ref.~\cite{usvyat-2011} lies closest to our result.

\begin{table}[H]
\caption{Total HF energies (in $\mathrm{Ha}$) for LiH and diamond obtained in this work and from the literature.
The CBL estimates given in the last column were obtained by extrapolation or perturbative correction, as specified in the footnotes.
}
\label{tab:lih_tot.vs.literature}
\centering
\begin{tabular}{llll}
\hline
\hline
\multicolumn{4}{c}{LiH}   \\
\hline
Source & Code/method & Variational & CBL estimate \\
\hline
This work                   & \texttt{exciting} (LAPW) & -8.064710 & --- \\
Sun \textit{et al.}  (2023) \cite{sun-2023} & PySCF (LCAO)             & -8.064539 & --- \\
Daga \textit{et al.}  (2020) \cite{daga-2020} & CRYSTAL (LCAO)          & -8.064618 & -8.065089\textsuperscript{\emph{a}}  \\
Usvyat \textit{et al.} (2011) \cite{usvyat-2011} & CRYSTAL (LCAO)          & -8.06460 & -8.06475\textsuperscript{\emph{b}}  \\
Civalleri \textit{et al.} (2010) \cite{civalleri-2010-paier-comment,Dovesi2026} & CRYSTAL (LCAO)       & -8.064544 & ---\\
Paier \textit{et al.} (2009) \cite{paier-2009}& CP2K (LCAO)            & -8.064545 &  ---\\
Paier \textit{et al.} (2009) \cite{paier-2009}& GAUSSIAN (LCAO)          & -8.064543\textsuperscript{\emph{c}} &  ---\\
\hline
\hline
\multicolumn{4}{c}{C}   \\
\hline
Source & Code/method & Variational & CBL estimate \\
\hline
This work & \texttt{exciting} (LAPW)& -75.775368 & --- \\
Daga \textit{et al.} (2020) \cite{daga-2020} & CRYSTAL (LCAO) & -75.774752 & -75.775983\textsuperscript{\emph{a}}\\
\hline
\hline
\end{tabular}

\textsuperscript{\emph{a}}~Two-data-point extrapolation;\\
\textsuperscript{\emph{b}}~Perturbative estimate of the total energy;\\
\textsuperscript{\emph{c}}~Extrapolated from the short-range to the bare-Coulomb kernel in the Fock exchange via a Padé fit.

\end{table}

Published cohesive energies cover a broader variety of methods, including LCAO~\cite{paier-2009,civalleri-2010-paier-comment,stoll-2012}, plane waves~\cite{Marsman2009,gruneis-2010}, and composite~\cite{gillan-2008} approaches.
Tab.~\ref{tab:coh-literature} compares our cohesive energies obtained for LiH and C with previously published data.
For LiH, four values from the literature agree with our result within 0.02~kcal/mol, while the full set of published data spans a few tenths of a kcal/mol. It is significant that such differences already appear for LiH, because LiH is the simplest periodic HF benchmark and extending comparable calculations to more complex solids is substantially more demanding.
Beyond those results in the literature, a further study
(see Tab.~VI in Ref.~\cite{gruneis-2010}) used the VASP code that relies on the projector-augmented waves (PAW) to calculate cohesive energies for seven materials from our set with the same geometry parameters as in Tab.~\ref{tab:solids}.
Their cohesive energy results per formula unit differ from ours by less than 1~kcal/mol for LiH, LiF, and C.
For MgO, Si, and BP, the discrepancy is 1--3~kcal/mol, and it reaches 16~kcal/mol for SiC.

\begin{table}[H]
\caption{Cohesive energies (in kcal/mol) obtained in this work and extracted from literature for LiH and diamond in the experimental geometry.}
\centering
\label{tab:coh-literature}
\begin{tabular}{llll}
\hline
\hline
Source & Code/method & \multicolumn{2}{c}{ $E_\mathrm{coh}$}\\
\cline{3-4}\\[-8pt]
 & & \multicolumn{1}{c}{LiH} & \multicolumn{1}{c}{C}\\[3pt]
\cline{1-4}\\[-8pt]
This work & \texttt{exciting} (LAPW)    & -82.806 & -121.702 \\
Daga \textit{et al.} (2020) ~\cite{daga-2020} & CRYSTAL (LCAO)            &  -83.0      & -122.0 \\
Stoll \textit{et al.} (2012) ~\cite{stoll-2012} & CRYSTAL (LCAO)              & -82.800 & ---\\
Civalleri \textit{et al.} (2010) ~\cite{civalleri-2010-paier-comment} & CRYSTAL (LCAO)        & -82.819 & ---\\
Grüneis \textit{et al.} (2010) ~\cite{gruneis-2010} & VASP (PAW)                & -82.6  & -121.8\\
Paier \textit{et al.} (2009) \cite{paier-2009}& CP2K (LCAO)\textsuperscript{\emph{a}}                   & -82.799 & ---\\
Marsman \textit{et al.} (2009) ~\cite{Marsman2009} & VASP (PAW)\textsuperscript{\emph{b}}                 & -82.63 & ---\\
Gillan \textit{et al.} (2008) \cite{gillan-2008} & PW+LCAO combined\textsuperscript{\emph{c}}             & -82.800 & ---\\
\hline
\hline
\end{tabular}

\textsuperscript{\emph{a}}~Energy obtained from three calculations (bulk LiH, and the atoms Li and H);\\
\textsuperscript{\emph{b}}~Extrapolated to an infinite $\mathbf{k}$-grid;\\
\textsuperscript{\emph{c}}~Used a composite approach employing plane waves and pseudopotentials in PWSCF and LCAO in all-electron nonperiodic code MOLPRO.\\
\end{table}

\subsection{Equation-of-state parameters}
For the cubic materials in our benchmark set, we also provide EOS parameters, including equilibrium lattice constants, bulk moduli, and their pressure derivatives.
To calculate them, we consider each crystal at 7 equidistant lattice constants ranging from about $-5$\% to $+$5\% of the equilibrium volume and fit to the Birch--Murnaghan EOS to the obtained total energies.
To converge the equilibrium lattice constant within 0.001~Å, we sample the Brillouin zone using $8\times 8\times 8$ $\mathbf{k}$-points for all materials, except for LiF, MgO, NaCl, and NaF where the $6\times 6\times 6$ grid is sufficient.
As a result, we obtain the EOS parameters as given in Tab.~\ref{tab:EOS}.

\begin{table}[H]
\caption{Equilibrium HF lattice constant $a_0$ (in \AA{}), bulk modulus $B_0$ (in GPa) and its derivative $B_0^\prime$ obtained from Birch–Murnaghan EOS fits for the studied crystalline solids.}
\label{tab:EOS}
\centering
\begin{tabular}{llrl}
\hline
\hline
Material & $a_0$ & \multicolumn{1}{c}{$B_0$} & $B_0^\prime$\\
\hline
LiH & 4.1096 & 32.04 & 3.32\\
LiF & 3.9944 & 77.77 & 4.26\\
BN  & 3.5993 & 429.76 & 3.49\\
C   & 3.5515 & 497.18 & 3.48\\
MgO & 4.1851 & 181.77 & 4.01\\
Si  & 5.5072 & 104.22 & 3.88\\
SiC & 4.3669 & 244.55 & 3.63\\
NaCl& 5.7764 & 22.27  & 4.68\\
AlP & 5.5411 & 94.88  & 3.79\\
MgS & 5.2868 & 76.83  & 3.97\\
NaF & 4.6211 & 51.46  & 4.65\\
BP  & 4.5834 & 175.26 & 3.53\\
\hline
\hline
\end{tabular}
\end{table}

Tab.~\ref{tab:eq-of-state-literaturee} compares our results with literature data for three materials: LiH, C, and Si. 
Along with the lattice constants and bulk moduli, we include cohesive energies computed at the equilibrium structures and do not discuss the pressure derivatives because they were not reported in most studies. 
Past studies report the LiH lattice constant and its bulk modulus of 4.101--4.111~\AA{} and 32.0--32.9~GPa, respectively.
The discrepancy among these results is relatively small, but it is not the case for the cohesive energies that are scattered within the range of 1.4~kcal/mol, i.e., more than the chemical accuracy threshold.
Our results are within the range of previously published data for all three quantities.
The cohesive energies reported by Gr\"{u}neis \textit{et al.} in Refs.~\cite{gruneis-2015,gruneis-2010} differ between themselves by 0.01~eV/atom. 
The discrepancy appears in the last given decimal and is likely a result of rounding.
Still, both values agree with our result within 1~kcal/mol.
For C and Si, we find differences from our lattice constants below 0.01~\AA{} in all calculations except for those from Ref.~\cite{McClain2017}. 
As in the LiH case, the cohesive energies are scattered by more than 1.0~kcal/mol.
For diamond, the GTH-HF pseudopotential and all-electron calculations of Ye \textit{et al.}~\cite{Berkelbach-2024}, performed with the same PySCF code and the corresponding Gaussian basis sets, yield cohesive energies differing by 3.3~kcal/mol among themselves. 
For all three materials, the PBE-derived LAPW setups change
$a_0$ and $B_0$ relative to the HF-consistent results by at most 0.0004~\AA{} and 0.12~GPa,
respectively.
Our cohesive-energy calculations use independently converged atomic reference energies and do not rely on error cancellation.
We therefore report cohesive energies only for the HF-consistent LAPW setups.

\begin{table}[H]
\caption{EOS parameters (equilibrium lattice constant $a_0$ in \AA{} and bulk modulus $B_0$ in GPa) and cohesive energy for the optimized lattice constant of the crystal (in kcal/mol), obtained in this work and extracted from the literature.}
\label{tab:eq-of-state-literaturee}
\centering
\begin{tabular}{lllll}
\hline
\hline
Source & Code/method & $a_0$  & $B_0$  & $E_{\mathrm{coh}}$ \\[3pt]
\cline{1-5}\\[-8pt] \\
\multicolumn{5}{c}{LiH}   \\
\cline{1-5}\\[-8pt]
This work & \texttt{exciting} (LAPW(--/HF))                                     & 4.1096 & 32.04 & -82.820 \\
This work & \texttt{exciting} (LAPW(--/PBE))   & 4.1095 & 32.03 &  ---\\
Ye \textit{et al.} (2022) \cite{Berkelbach-2022} & PySCF (LCAO)                     & 4.101 & 32.6 & -83.5 \\
Grüneis \textit{et al.} (2015) \cite{gruneis-2015} & VASP (PAW/PW)                       & 4.111 & 32.4 & -82.1 \\
Grüneis \textit{et al.} (2010) \cite{gruneis-2010} & VASP (PAW/PW)              & 4.111 & 32 & -82.6 \\
Civalleri \textit{et al.} (2010) \cite{civalleri-2010-paier-comment}& CRYSTAL (LCAO)       & 4.105 & 32.9 &  ---\\
Paier \textit{et al.} (2009) \cite{paier-2009}& GAUSSIAN (LCAO)                        & 4.105  & 32.34 &   ---\\
Gillan \textit{et al.} (2008) \cite{gillan-2008}& PW+LCAO combined\textsuperscript{\emph{a}} & 4.1084 & 32.05 & -82.825 \\
\\
\multicolumn{5}{c}{C}   \\
\cline{1-5}\\[-8pt]
This work & \texttt{exciting} (LAPW(--/HF))                                     & 3.5515 & 497.18 & -121.736 \\
This work & \texttt{exciting} (LAPW(--/PBE))     & 3.5516 & 497.06 &  ---\\
Ye \textit{et al.} (2024) \cite{Berkelbach-2024} & PySCF (LCAO/GTH)                 & 3.557 & 490 & -119.2 \\
Ye \textit{et al.} (2024) \cite{Berkelbach-2024} & PySCF (LCAO/All-e)               & 3.550 & 498 & -122.5 \\
McClain \textit{et al.} (2017) \cite{McClain2017} & PySCF (LCAO/GTH)                        & 3.527 & 507 & -123.6 \\
Grüneis \textit{et al.} (2010) \cite{gruneis-2010} & VASP (PAW/PW)                       & 3.552 & 495 & -121.8 \\
\\
\multicolumn{5}{c}{Si}   \\
\cline{1-5}\\[-8pt]
This work & \texttt{exciting} (LAPW(HF/HF))                                    & 5.5072 & 104.22 & -70.500 \\
This work & \texttt{exciting} (LAPW(PBE/PBE))    & 5.5076 & 104.17 &  ---\\
McClain \textit{et al.} (2017) \cite{McClain2017} & PySCF (LCAO/GTH)                         & 5.435 & 107 & -69.9 \\
Grüneis \textit{et al.} (2010) \cite{gruneis-2010}  & VASP (PAW/PW)                        & 5.512 & 103 & -68.5 \\
\hline
\hline
\end{tabular}

\textsuperscript{\emph{a}}~Used a composite approach employing plane waves and pseudopotentials in PWSCF and LCAO in the all-electron nonperiodic code MOLPRO.
\end{table}

\subsection{Error cancellation in crystal formation energies}
As in the molecular case, relative energies in solids also benefit substantially from error cancellation.
To illustrate this point, we assess how sensitive energy differences are in solids to the following settings: (i) the LAPW cutoff $R_\mathrm{MT}K_\mathrm{max}$, (ii) HF- versus PBE-derived radial basis functions, (iii) HF- versus PBE-derived core orbitals.
For this purpose, we compute the formation energy of SiC defined as $E_\mathrm{form}=E_\mathrm{bulk}(\mathrm{SiC})-E_\mathrm{bulk}(\mathrm{Si})-E_\mathrm{bulk}(\mathrm{C})$, where $E_\mathrm{bulk}(\mathrm{X})$ is the bulk total energy per formula unit of a solid X using the same lattice parameters as in Tab.~\ref{tab:solids}.

Tab.~\ref{tab:SiC-formation} presents results of this assessment as a hierarchy of simplifications in the computations. 
The first result corresponds to the reference based on the total energies reported in Tab.~\ref{tab:solids}.
Reducing the LAPW cutoff to $R_\mathrm{MT}^{\mathrm{C}}K_\mathrm{max}=8$ alters the formation energy by only $0.004~\mathrm{kcal/mol}$.
At the reduced cutoff, replacing the HF-derived radial basis functions by PBE-derived ones changes the result by a further $0.006~\mathrm{kcal/mol}$, whereas additionally using PBE-derived core orbitals changes it by less than $10^{-4}~\mathrm{kcal/mol}$.
Thus, the strong error cancellation found for molecular formation energies is
also observed for the SiC formation energy in the solid state.

\begin{table}[H]
\caption{\label{tab:SiC-formation}
Formation energy (in kcal/mol) of SiC obtained with converged and reduced plane-wave basis size, with different basis-function and core-orbital types.}
\centering
\begin{tabular}{ccl}
\hline
\hline
 $R^{\mathrm{C}}_\mathrm{MT}K_\mathrm{max}$  & Calculation setup & $E_{\mathrm{form}}$  \\
 \hline
11 & LAPW(HF/HF)  & -17.7178 \\
8 & LAPW(HF/HF)   & -17.7142 \\
8 & LAPW(HF/PBE)   & -17.7085 \\
8 & LAPW(PBE/PBE) & -17.7086 \\
\hline
\hline
\end{tabular}
\end{table}

\subsection{Self-interstitials} 
As an illustration of the practical use of the presented method, we compute the HF contribution to the formation energies of Si self-interstitials in the hexagonal (H) and split (X) configurations, which have recently been studied with post-HF approaches~\cite{Gruneis-2023-Si-defects,Simula-Si-defects-2026}.
These two studies explored several electron-correlation methods, yet reproducibility becomes an issue already at the HF level of theory, as different families of basis sets and pseudopotentials yield different results despite Si being one of the most studied chemical elements in the solid state.

To assess past calculations and to provide a reference, we use 17-atom supercells with the same structures as in Refs.~\cite{Gruneis-2023-Si-defects,Simula-Si-defects-2026} and calculate the defect formation energies defined as
\begin{equation}
E^{\mathrm{form}} = E^{\mathrm{def}} - \frac{N_{\mathrm{def}}}{N_{\mathrm{bulk}}} E^{\mathrm{bulk}},
\end{equation}
with $N_{\mathrm{def}}=17$ and $N_{\mathrm{bulk}}=2$.
We set the LAPW cutoff $R^{\mathrm{Si}}_\mathrm{MT}K_\mathrm{max}=12$ with $R^{\mathrm{Si}}_\mathrm{MT}=2.0~a_\mathrm{B}$ to ensure high precision for the defect formation energies.
The Brillouin-zone sampling is performed using a $\Gamma$-centered
$4 \times 4 \times 4$ $\mathbf{k}$-point grid.
We estimate the correction for a denser $5 \times 5 \times 5$ $\mathbf{k}$-point grid from computationally cheaper tests in which the $1s^2 2s^2 2p^6$ shells are treated as core states. 
The corrections amount to only 0.10 and 0.04~kcal/mol for the H and X formation energies, respectively, and are included in our final results, where only the $1s$ state is treated as core.

\begin{figure}[H]
    \centering
    \includegraphics[width=280pt]{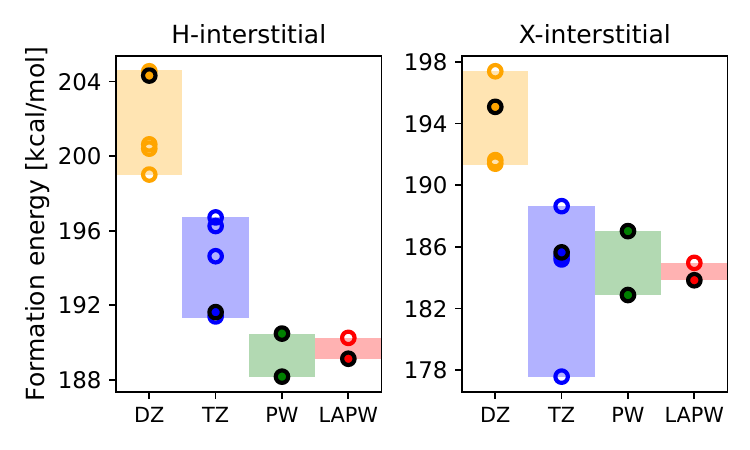}
    \caption{
    Formation energies (in kcal/mol) of Si self-interstitials in the hexagonal (H) and split (X) configurations obtained in this work and reported in the literature. The literature data include all-electron and pseudopotential LCAO calculations with CRYSTAL (labeled with DZ and TZ corresponding to double- and triple-$\zeta$, respectively) and plane-wave pseudopotential calculations with Quantum ESPRESSO reported in Ref.~\cite{Simula-Si-defects-2026} and VASP calculation reported in Ref.~\cite{Gruneis-2023-Si-defects}. Black markers denote calculations that include scalar-relativistic effects. All other markers correspond to nonrelativistic results.
    }
    \label{fig:si-interstitials}
\end{figure}

Fig.~\ref{fig:si-interstitials} compares our results with previous nonrelativistic all-electron and scalar-relativistic pseudopotential calculations.
To provide a reference for both cases, we perform two types of calculations: (i) using the nonrelativistic Hamiltonian, and (ii) employing the atomic zeroth-order regular approximation (aZORA)~\cite{VanLenthe1993,VanLenthe1994}.
aZORA uses a fixed atomic potential inside the atomic spheres and avoids the gauge-dependent energy drift of conventional ZORA.
The Si $1s$ core state is treated using the four-component Dirac equation with a local
PBE potential.
We find the LAPW(Dirac PBE/aZORA HF) setup sufficiently reliable in this application, as replacing the Dirac PBE core by an aZORA HF core changes the H-defect formation energy by only $0.05~\mathrm{kcal/mol}$.
In comparison to the nonrelativistic LAPW(HF/HF) result calculated using $R^{\mathrm{Si}}_\mathrm{MT}K_\mathrm{max}=12$, reducing the cutoff to $R^{\mathrm{Si}}_\mathrm{MT}K_\mathrm{max}=8$ and employing the LAPW(PBE/PBE) setup changes the formation energies by only 0.08 and 0.11~kcal/mol for the H and X defects, respectively.
These results closely match the reference values and are not shown separately in Fig.~\ref{fig:si-interstitials}. They are reported in Tab.~S3 of the Supporting Information.
Additional tests assessing the relativistic treatment, together with a detailed
comparison to literature results, are provided in the Supporting Information.
Our final nonrelativistic formation energies (including the $\mathbf{k}$-point grid correction) are 190.25 and 184.96~kcal/mol for the H and X self-interstitials, respectively, yielding an H-X formation-energy difference of $\Delta E_{\mathrm{H}-\mathrm{X}}=5.29$~kcal/mol.
Including relativistic effects yields 189.14 and 183.77~kcal/mol, respectively, while leaving the relative energy nearly unchanged, $\Delta E_{\mathrm{H}-\mathrm{X}}=5.36$~kcal/mol.
Relativistic effects shift the absolute defect formation energies by approximately $1~\mathrm{kcal/mol}$ but have little effect on the H--X energy difference that determines the relative thermodynamic stability of the two configurations.

Fig.~\ref{fig:si-interstitials} provides the assessment of the literature data and allows us to draw conclusions on the performance and limitations of the computational setups used in Refs.~\cite{Gruneis-2023-Si-defects,Simula-Si-defects-2026}.
All-electron GTO calculations improve the defect formation energies when a double-$\zeta$ basis is replaced by its triple-$\zeta$ counterpart within the same basis set family. 
Even the triple-$\zeta$ outlier underestimating the formation energy of the X defect at 177.6~kcal/mol provides an improvement over the corresponding double-$\zeta$ result of 197.4~kcal/mol, although the H--X energy difference increases from 7.1 to 19.1~kcal/mol, moving further from the LAPW reference of 5.29~kcal/mol (see Tab.~S3 of the Supporting Information for additional details).
Among the all-electron triple-$\zeta$ results, the best agreement with our non-relativistic calculation is obtained for the def2-TZVP~\cite{weigend2005a} basis set with the deviation of only 1.2~kcal/mol and 0.2~kcal/mol for the H- and X-defects, respectively.
Although such a small difference seems encouraging, the ranges of the energies obtained with the different triple-$\zeta$ basis sets is 5.3~kcal/mol and 11.0~kcal/mol depending on the interstitial configuration.
Even if cherry-picking the best result from these data with having a reference may be possible based on a prior experience and other arguments, there is no guarantee that the selected result is sufficiently precise.
The next step is clear when studying molecules -- one simply runs quadruple and possibly quintuple-$\zeta$ calculations as in the example shown in Fig.~\ref{fig:molecule-reaction-errors}.
In solid-state systems, however, such a progression is typically not available for GTOs due to issues with numerical linear dependence. 
It is possible to refine the energies within the GTO framework using perturbative approaches~\cite{Wolinski_2003_MP2_dual-basis,Maschio2024}, but their precision requires further quantification.

The GTO+pseudopotential calculations have the same issues with high-$\zeta$ basis sets, however, the energies can be verified using a plane-wave code that yields the complete-basis result (loosely speaking, $\infty\zeta$) for the given pseudopotential (ccECP~\cite{ccECP}).
Such an improvement (triple-$\zeta\rightarrow\infty\zeta$) leads to corrections of 1.1~kcal/mol and 1.4~kcal/mol for the H- and X-defects, respectively.
However, the energies obtained in plane-wave calculations differ from our scalar-relativistic LAPW results by larger gaps of 1.4~kcal/mol and 3.2~kcal/mol.
Another plane-wave calculation employed projector-augmented waves (or simply PAW pseudopotentials) and obtained the defect formation energies that differ from our values and the plane-wave ccECP results by 0.9--1.0~kcal/mol and 2.3--4.1~kcal/mol, respectively. 
This point serves as a reminder that approximations introduced by pseudopotentials are also a source of errors.
Performance of pseudopotentials depends on details of their parametrization, such as which atomic properties were targeted and with what weights, which Hamiltonian (for example, DFT versus HF) was used, and what radial constraints were enforced. 
The individual influence of each detail is hard to predict, but the overall performance of pseudopotentials can now be assessed reliably also for HF calculations.

\subsection{Band Gaps and Core-Level Transitions}

The reference values for the band gaps from HF calculations are given in Tab.~\ref{tab:solids}, and they are obtained as a byproduct of the total-energy data.
These data are useful mainly for verifying implementations, basis quality and pseudopotentials, whereas a comparison to experiment rather requires hybrid exchange-correlation functionals.
For this reason, we continue our analysis using PBE0~\cite{Perdew1996-hyb} for the NaCl crystal using the same geometry as in Tab.~\ref{tab:solids}.

A nonrelativistic LAPW(PBE0/PBE0) calculation with $R_\mathrm{MT}^{\mathrm{Na}} =R_\mathrm{MT}^{\mathrm{Cl}}=1.4~a_\mathrm{B}$, $R_\mathrm{MT}K_\mathrm{max}=12$ and $10\times10\times10$ $\mathbf{k}$-points yields a band gap of 7.3865~eV. 
Based on our numerical tests with respect to the basis and $\mathbf{k}$-points settings, we estimate the numerical uncertainty of this result as 0.1~meV.
We note that the chosen muffin-tin radii are small for this material, and it is more common to set them to such values that $R_\mathrm{MT}^{\mathrm{Na}} + R_\mathrm{MT}^{\mathrm{Cl}}$ approaches the interatomic distance of $5.28~a_\mathrm{B}$.
Larger muffin-tins result in a smaller number of LAPWs required to ensure the same level of precision in the interstitial region, thus, leading to cheaper calculations.
Increasing $R_\mathrm{MT}^{\mathrm{Na,Cl}}$ without a loss of precision may, however, require to increase the radial basis.
Below, we verify whether this is necessary in the LAPW(PBE0/PBE0) and LAPW(PBE/PBE) cases.

\begin{figure}[H]
  \centering
  \includegraphics[width=470pt]{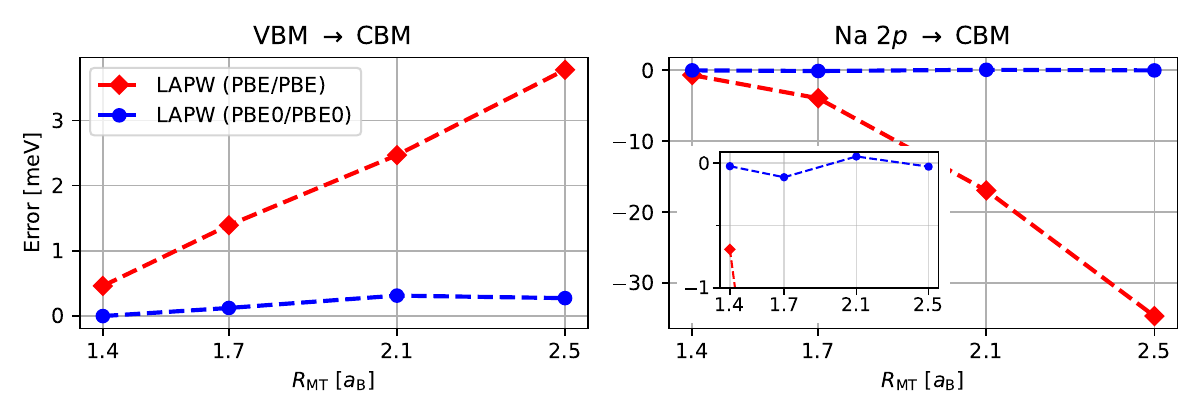}
  \caption{\label{fig:bandgap}
    Deviation of the PBE0 transition energies (the valence band maximum to the conduction band minimum on the left and Na $2p$ to the conduction band minimum on the right) in NaCl with respect to the reference value as a function of the muffin-tin radius $R_\mathrm{MT}^{\mathrm{Na,Cl}}$.
  In the LAPW(PBE/PBE) approach, both the core orbitals and radial basis functions are generated at the PBE level, whereas in the LAPW(PBE0/PBE0) case, both are generated consistently using PBE0.
  Deviations are given relative to the LAPW(PBE0/PBE0) result obtained with $R_\mathrm{MT}=1.4~a_\mathrm{B}$ and additional radial basis functions.
  }
\end{figure}

The left panel of Figure~\ref{fig:bandgap} compares the band gaps obtained setting $R_\mathrm{MT}^{\mathrm{Na,Cl}}=1.4$--$2.5~a_\mathrm{B}$.
We find that the band gap calculated using the LAPW(PBE0/PBE0) approach remains within a sub-meV range.
All these calculations employ the same number of radial basis functions (for Na and for Cl), and the product $R_\mathrm{MT}K_\mathrm{max}$ remains fixed as the muffin-tin radius changes.
The same computational setup in the LAPW(PBE/PBE) case leads to a slightly stronger $R_\mathrm{MT}^{\mathrm{Na,Cl}}$ dependence.
The band gap calculated at $R_\mathrm{MT}^{\mathrm{Na,Cl}}=1.4~a_\mathrm{B}$ reproduces the reference result with the deviation of 0.5~meV.
Increasing $R_\mathrm{MT}^{\mathrm{Na,Cl}}$ leads to an increase of the error reaching 3.8~meV at 2.5~$a_\mathrm{B}$.
Finally, we note that the main source of errors in this calculation are specifically PBE-derived radial basis function, whereas switching the PBE and PBE0 cores ($1s^2$ shells in Na and Cl) altered the band gap by less than 0.1~meV. 

The maximum deviation of 3.8~meV found for NaCl is consistent with the error estimates of 10~meV for GaAs and 6~meV for ZnSe in Refs.~\cite{Zavickis_ACE} and \cite{Uzulis2025}, respectively.
The present results thus support the earlier assessment that errors associated with the PBE-derived construction remain small compared with the typical target precision of 50--100~meV in practical band-gap calculations.

We also consider the independent-particle transition from the Na $2p$ core level to the conduction-band minimum (CBM).
The transition energy calculated as the difference of the corresponding orbital eigenvalues provides an
independent-particle starting point for describing the
Na $L_{2,3}$ absorption edge.

We limit our discussion to numerical errors, as resolving in detail all features of the absorption spectra requires including at least the spin-orbit interaction and a many-body treatment as in the Bethe-Salpeter equation with an input from a DFT calculation directly or with a $GW$ correction of orbital energies in an additional intermediate step.

The right panel of Fig.~\ref{fig:bandgap} reveals a substantially stronger dependence on the muffin-tin radius for this transition in the LAPW(PBE/PBE) setup.
At $R_\mathrm{MT}^{\mathrm{Na,Cl}}=2.5~a_\mathrm{B}$, the transition energy is underestimated by approximately 35~meV, compared with an overestimation of approximately 3.8~meV for the fundamental gap.
The error mainly stems from the radial functions, as replacing the PBE core with the ones obtained from PBE0 leads to a small further increase of the error by 1.3~meV at $R_\mathrm{MT}^{\mathrm{Na,Cl}}=2.5~a_\mathrm{B}$.
In contrast, the LAPW(PBE0/PBE0) deviations remain within approximately 0.2~meV of the reference throughout the radius range considered.
This agreement is maintained without increasing the number of radial basis functions.
The functional-consistent construction substantially maintains the numerical precision of this transition at larger muffin-tin radii, where fewer LAPWs are required.

\section{CONCLUSIONS} 
In conclusion, we have removed a fundamental limitation of periodic Hartree--Fock calculations in the LAPW framework by constructing radial basis functions and core orbitals consistently with the HF Hamiltonian. This development enables direct all-electron HF reference calculations with near-complete-basis precision for both molecules and solids. We have validated the approach against molecular benchmark data and established reference-quality HF total and cohesive energies for a set of semiconductors and insulators. For the examples studied here, the standard simplified approach based on PBE-derived radial functions and core orbitals introduces only small errors in practical relative energies.
For the radial bases and core--valence partitions investigated here, HF-consistent radial functions and core orbitals provide a practical route to converged absolute HF energies.

The proposed method and benchmarks offer a stringent validation target for periodic all-electron methods such as finite elements and, once extended to periodic boundary conditions, multiwavelets. Our data also provide a benchmark for LCAO basis-set development and for assessing pseudopotential-based calculations. Since the construction is not restricted to HF, it is also directly applicable to hybrid density functionals within LAPW, thereby extending the present approach to a broader class of chemically and physically relevant calculations. The explicit access to core orbitals further enables the use of hybrid functionals within LAPW for modeling core-level spectroscopy and provides improved starting points for many-body methods, including $GW$ and the Bethe--Salpeter equation.

The present work also hints at the necessary future extensions: spin-polarized calculations, forces, and a broader selection of relativistic approximations, especially those including the spin-orbit coupling.
Once the first two of the listed features are implemented even with the PBE-derived radial functions and core orbitals, their extension to the current approach with the consistent basis is straightforward.
The relativistic extensions, in turn, require solving the Dirac equation for the core electrons and possibly some approximation to it
for the valence radial functions.
This can be achieved following a similar strategy to what is applied in the current study, although the implementation details will need to be adjusted.

\section*{DATA AND SOFTWARE AVAILABILITY}
All input and output data of the calculations performed in this work are available in the NOMAD repository~\cite{Nomad_dataset}.
Additional computational details and convergence tests are provided in the Supporting Information.
The modified \texttt{exciting} source code implementing the method described in this work is publicly available on Github at
\url{https://github.com/gulans/exciting/tree/ace-core}~.\\

\section*{SUPPORTING INFORMATION}

The Supporting Information file SI.pdf is available free of charge and contains:
\begin{itemize}
    \item Cl$_2$ total-energy convergence test with respect to the cutoff parameter $R^{\mathrm{Cl}}_\mathrm{MT}K_\mathrm{max}$.
    \item Total-energy convergence for the diamond crystal using PBE- and HF-derived radial basis functions in the $s$ and $p$ channels.
    \item Formation energies of FH, ClF, and ClF$_5$ calculated using various LAPW basis-set configurations and GTO basis sets.
    \item Convergence behavior of the total energy for crystalline systems with respect to $\mathbf{k}$-point sampling.
    \item Formation energies of Si self-interstitial defects, along with a summary of computational details.
\end{itemize}

\section*{ACKNOWLEDGMENTS}
This research is funded by the Latvian Council of Science, project "Precise first principles methods for modelling quantum materials", project No. lzp-2024/1-0202.
Jānis Užulis acknowledges funding provided by the project “Strengthening the Research and Development Capacity of Doctoral Studies at the University of Latvia in the Fields of Smart Specialisation”, identification no. 1.1.1.8/1/24/I/003.
We acknowledge EuroHPC JU for awarding the project ID EHPC-DEV-2025D04-071 access to the EuroHPC supercomputer MareNostrum 5 (hosted by the Barcelona Supercomputing Centre) through a EuroHPC Development Access call.
Finally, we are grateful to Denis Usvyat for pointing out the Si self-interstitial problem.

\printbibliography

\newpage
\section*{TOC GRAPHIC}
\begin{figure}[H]
    \includegraphics[width=3.25in,height=1.75in]{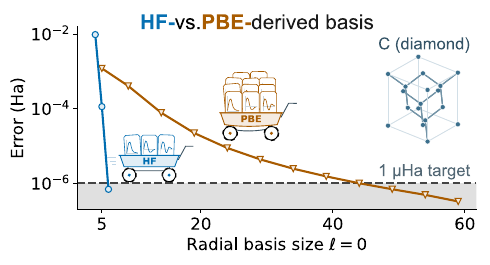}
\end{figure}

\newpage

\input{supp-mat-arxiv}

\end{document}

%% file: supp-mat-arxiv.tex
\setcounter{figure}{0}
\renewcommand{\thefigure}{S\arabic{figure}}
\setcounter{table}{0}
\renewcommand{\thetable}{S\arabic{table}}
\setcounter{equation}{0}
\renewcommand{\theequation}{S\arabic{equation}}


\textbf{\Large Supporting Information: Hartree--Fock Limit for Energies in Solids}


\section{Basis convergence: computational details and results}

\begin{figure}[H]
    \centering
    \includegraphics[width=240pt]{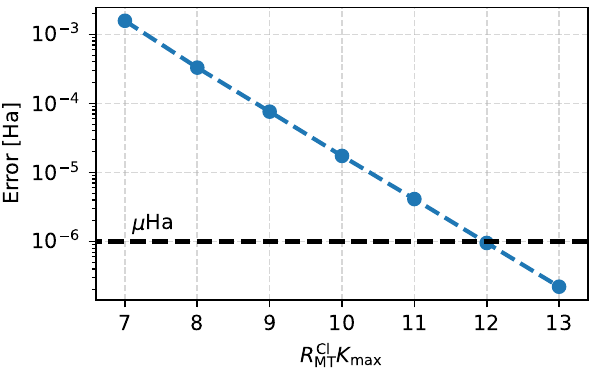}
    \caption{Error in the total HF energy for the Cl$_2$ molecule with respect to $R^\mathrm{Cl}_\mathrm{MT}K_\mathrm{max}$. 
  The errors are estimated using the result obtained with $R^\mathrm{Cl}_\mathrm{MT}K_\mathrm{max}=14$ as the reference.}
    \label{fig:rgkmax_conv}
\end{figure}

\begin{figure}[H]
    \centering
    \includegraphics[width=360pt]{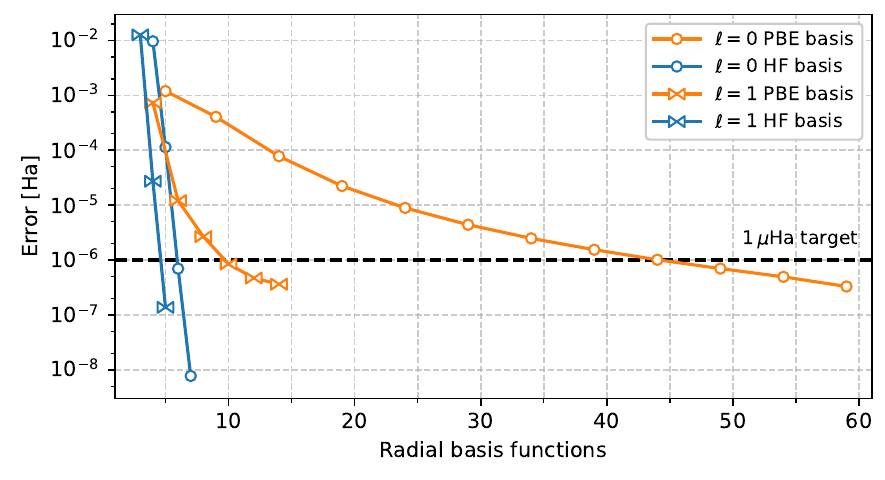}
    \caption{
    Error in the total HF energy (in Ha per unit cell) of diamond with respect to the number of radial basis functions in the $s$ ($\ell=0$) or $p$ ($\ell=1$) channel, comparing PBE- and HF-derived radial basis functions. The errors are evaluated relative to the result obtained with 9 and 7 radial basis functions in the $s$ and $p$ channels, respectively. Lines connecting the data points are guides to the eye.}
    \label{fig:diamond_lo_conv}
\end{figure}

To attain the target precision of a few $\mu$Ha, we converge the total energy with respect to (i) the number of plane waves used for describing wave functions in the interstitial region and (ii) the number of high-energy radial basis functions.
The item (i) is controlled by the dimensionless cutoff parameter $R_\mathrm{MT}K_\mathrm{max}$. 
Fig.~\ref{fig:rgkmax_conv} shows the convergence of the total energy for Cl$_2$ with respect to $R^\mathrm{Cl}_\mathrm{MT}K_\mathrm{max}$. The observed convergence is consistent with the trends observed in Refs.~\cite{Gulans2014,Gulans2018,Zavickis_ACE}.
The item (ii) is equivalent to adding local orbitals following the terminology commonly accepted in the LAPW community and it is discussed in the main text of the paper for Cl$_2$ molecule.
To compare the convergence obtained with PBE- and HF-derived
radial basis functions, we consider the diamond crystal.
Fig.~\ref{fig:diamond_lo_conv} shows the total HF energy error
as a function of the number of radial basis functions in the
$s$ and $p$ channels. The HF-derived basis yields much faster
convergence in both $\ell$ channels.

\section{Molecular formation energies: computational details and results}

Tab.~\ref{tab:molec-reactions} reports the formation energies for FH, ClF, and ClF$_5$ with different LAPW and GTO basis sets.
When molecular formation energies are computed with the reduced $R_\mathrm{MT}K_\mathrm{max}$ values, error cancellation requires consistent computational settings, i.e., the same muffin-tin radii and same radial-basis sets for each species.

\begin{table}[H]
\caption{\label{tab:molec-reactions}Molecular formation energies from HF calculations carried out using the HelFEM, \texttt{exciting}, and NWChem codes. Energies whose differences from the reference result exceed 1~kcal/mol are shown in bold.}
\centering
\begin{tabular}{llll}   
\hline
\hline
Basis  &\multicolumn{3}{c}{  Formation energy [kcal/mol] }\\[3pt]
\cline{2-4}\\[-8pt]
         & FH\textsuperscript{\emph{a}} & ClF     & ClF$_5$    \\
\cline{1-4}\\[-8pt]
\multicolumn{4}{c}{\textsc{HelFEM}\textsuperscript{\emph{b}} }\\[3pt]
                             & -73.587\textsuperscript{\emph{c}}  & -16.589 & -        \\
\cline{1-4}\\[-8pt]
\multicolumn{4}{c}{\texttt{exciting}\textsuperscript{\emph{d}} (HF core / HF basis)} \\[3pt]
$R^{\mathrm{F}}_\mathrm{MT}K_\mathrm{max}=13$  & -73.587  & -16.591 & -32.650 \\
$R^{\mathrm{F}}_\mathrm{MT}K_\mathrm{max}=10$  & -73.587 & -16.592 & -32.650 \\
$R^{\mathrm{F}}_\mathrm{MT}K_\mathrm{max}=9$   & -73.586  & -16.593 & -32.651 \\
$R^{\mathrm{F}}_\mathrm{MT}K_\mathrm{max}=8$   & -73.581  & -16.595 & -32.648 \\
\cline{1-4}\\[-8pt]
\multicolumn{4}{c}{\texttt{exciting}\textsuperscript{\emph{d}} (HF core / PBE basis)} \\[3pt]
$R^{\mathrm{F}}_\mathrm{MT}K_\mathrm{max}=8$  & -73.589 & -16.593 & -32.634 \\
\cline{1-4}\\[-8pt]
\multicolumn{4}{c}{\texttt{exciting}\textsuperscript{\emph{d}} (PBE core / PBE basis)} \\[3pt]
$R^{\mathrm{F}}_\mathrm{MT}K_\mathrm{max}=8$  & -73.589   &  -16.591 & -32.613 \\

\cline{1-4}\\[-8pt]
\multicolumn{4}{c}{NWChem aug-cc-pwCV\textit{X}Z\textsuperscript{\emph{d}}}             \\[3pt]
aug-cc-pwCV5Z & -73.586 & -16.588 & -32.658 \\
aug-cc-pwCVQZ & -73.622 & -16.630 & -32.987 \\
aug-cc-pwCVTZ & -73.501 & -16.288 & -31.900 \\
aug-cc-pwCVDZ & \bf -75.080 & \bf -17.858 &\bf -19.610 \\

\cline{1-4}\\[-8pt]
\multicolumn{4}{c}{NWChem aug-cc-pCV\textit{X}Z\textsuperscript{\emph{d}}}             \\[3pt]
aug-cc-pCV5Z & -73.586 & -16.587 & -32.655 \\
aug-cc-pCVQZ & -73.625 & -16.624 & -32.904 \\
aug-cc-pCVTZ & -73.515 & -16.249 & \bf -30.118 \\
aug-cc-pCVDZ & \bf -75.104 & -17.334 &\bf -5.166 \\

\cline{1-4}\\[-8pt]
\multicolumn{4}{c}{NWChem aug-cc-pV\textit{X}Z\textsuperscript{\emph{d}}}             \\[3pt]
aug-cc-pV6Z  & -73.586  & -16.566 & -32.045 \\
aug-cc-pV5Z  & -73.587  & -16.509 &\bf -30.928 \\
aug-cc-pVQZ  & -73.623  & -16.354 &\bf -25.671 \\
aug-cc-pVTZ  & -73.522  & -15.849 &\bf -19.553 \\
aug-cc-pVDZ  &\bf -75.095  & -17.128 &\bf -0.178 \\

\hline
\hline
\end{tabular}

\textsuperscript{\emph{a}}~FH contains no core orbitals in the \texttt{exciting} calculations. For the Cl-containing molecules, the Cl 1s orbital is the only core orbital;\\
\textsuperscript{\emph{b}}~Extracted from Ref.~\cite{Lehtola2019} if not noted otherwise;\\
\textsuperscript{\emph{c}}~For the FH formation energy, the $\mathrm{H}_2$ total energy is obtained in the present work, whereas the remaining molecular energies are taken from Ref.~\cite{Lehtola2019};\\
\textsuperscript{\emph{d}}~Present work.
\end{table}

\section{Total energies in solids: convergence with respect to the \textbf{k}-grid}
To converge the total energy with respect to the $\mathbf{k}$-grid, we perform calculations on $N \times N \times N$ $\Gamma$-centered grids increasing $N$ in steps of two.
Comparing two consecutive calculations, we find that the error in the total energy decreases by at least a factor of 8 with each step. 
For each material, we consider the total energy converged with respect to $\mathbf{k}$-point sampling when the absolute difference between the results obtained using the two densest meshes considered for that material is less than 7~$\mu$Ha per unit cell.
Tab.~\ref{tab:k-convergence} lists the total energies obtained for all considered $\mathbf{k}$-point meshes.
\begin{table}[H]
  \caption{Total energies (in Ha) obtained for the studied materials depending on the $\mathbf{k}$-point sampling.
}
  \label{tab:k-convergence}
  \centering
  \begin{tabular}{lrrrrrrr}
    \hline
    Material  & $2 \times 2 \times 2$ & $4 \times 4 \times 4$ & $6 \times 6 \times 6$ & $8 \times 8 \times 8$ & $10 \times 10 \times 10$ & $12 \times 12 \times 12$ \\
    \hline
LiH (RS) & -8.0376562 & -8.0642578 & -8.0646885 & -8.0647089 & -8.0647103 & -8.0647104 \\
LiF (RS) & -107.0815930 & -107.0942393 & -107.0942758 & -107.0942760 &  --- &  --- \\
BeO (WUR)\textsuperscript{\emph{a}} & -179.4387470 & -179.4769278 & -179.4773687 & -179.4773789 & -179.4773791 & ---  \\
BN (ZB) & -79.1803916 & -79.2820364 & -79.2845054 & -79.2846013 & -79.2846058 &  --- \\
BP (ZB) & -365.3895988 & -365.4988497 & -365.5035437 & -365.5038637 & -365.5038895 & -365.5038918 \\
C (DIA) & -75.6345501 & -75.7704488 & -75.7750952 & -75.7753501 & -75.7753669 & -75.7753682 \\
NaF (RS) & -261.4828270 & -261.4870488 & -261.4870536 & ---  &  --- &  --- \\
NaCl (RS) & -621.5257195 & -621.5361724 & -621.5362139 & -621.5362141 &  --- & ---  \\
MgO (RS) & -274.6697317 & -274.7014582 & -274.7017264 & -274.7017303 &  --- &  --- \\
MgS (RS) & -597.2810574 & -597.3321133 & -597.3332029 & -597.3332396 & -597.3332413 & ---  \\
AlN (WUR)\textsuperscript{\emph{b}} & ---  & -593.1473985 & -593.1480719 & -593.1480880 & -593.1480884 &  --- \\
AlP (ZB) & -582.7340765 & -582.7988994 & -582.8010195 & -582.8011272 & -582.8011338 & ---  \\
SiC (ZB) & -326.7877723 & -326.8832806 & -326.8865576 & -326.8867349 & -326.8867465 & -326.8867472 \\
Si (DIA) & -577.8516781 & -577.9373551 & -577.9413361 & -577.9416280 & -577.9416528 & -577.9416559 \\
    \hline
  \end{tabular}
  \textsuperscript{\emph{a}} The values for BeO correspond to the \textbf{k}-grid of $2 \times 2 \times 1$, $4 \times 4 \times 2$, $6 \times 6 \times 3$,	$8 \times 8 \times 4$, and $10 \times 10 \times 6$;\\
  \textsuperscript{\emph{b}} The values for AlN correspond to \textbf{k}-grid of $4 \times 4 \times 3$, $6 \times 6 \times 4$, $8 \times 8 \times 5$, and $10 \times 10 \times 6$.
\end{table}

\section{Self-interstitials in Si: computational details, results, and literature review}

Tab.~\ref{tab:si-interstitials} summarizes our LAPW formation energies together with literature results obtained using Gaussian type orbitals and plane-wave (PW) basis sets.
These literature data and our final LAPW results are plotted in Fig.~3 of the main text.

We also discuss how the formation energies are influenced by the standard LAPW(PBE/PBE) setup and a choice of relativistic treatment.
The results for molecular and crystal formation energies in the main text show that error cancellation keeps the errors from the standard LAPW(PBE/PBE) setup small.
To assess whether this also holds for Si self-interstitials, we repeat the nonrelativistic HF calculations using LAPW(PBE/PBE) with a reduced cutoff of $R^{\mathrm{Si}}_\mathrm{MT}K_\mathrm{max}=8$.
Compared with the nonrelativistic LAPW(HF/HF) reference calculations at $R^{\mathrm{Si}}_\mathrm{MT}K_\mathrm{max}=12$, the formation energies differ by only 0.08 and 0.11~kcal/mol for the H and X defects, respectively.
Thus, the standard LAPW(PBE/PBE) approach introduces only a small error in these formation energies, even with the reduced cutoff.

Our final relativistic results in the main text use the LAPW(Dirac PBE/aZORA HF) core and basis function setup.
Ideally, the core orbitals would be generated using the Dirac-Fock equation, denoted LAPW(Dirac HF/HF), but this setup is not yet implemented in the code.
The four-component Dirac equation for core electrons is currently available only for (semi)local exchange-correlation functionals such as PBE. 
Scalar-relativistic effects in the valence Hamiltonian can likewise be treated only within ZORA and aZORA, while infinite-order relativistic approximations is currently available only for (semi)local exchange-correlation functionals.
It is therefore necessary to assess the reliability of the chosen setup and estimate the spread in formation energies arising from the available relativistic models.

To this end, we consider a large-core configuration ($1s^2 2s^2 2p^6$), whereas only the Si $1s$ shell is treated as core in the main calculations.
In the relativistic large-core calculation, all three core shells are described by the Dirac-Kohn-Sham equation with a local PBE potential.
This choice also reduces the computational cost by approximately an order of magnitude.
Specifically, we compare LAPW($1s^2 2s^2 2p^6$ Dirac PBE/ZORA PBE), which combines a Dirac core with a ZORA valence treatment, with LAPW($1s^2 2s^2 2p^6$ NR PBE/NR PBE), where NR denotes a nonrelativistic treatment.
Subtracting the nonrelativistic formation energies from the relativistic ones yields corrections of $-1.08$ and $-1.13$~kcal/mol for the H and X defects, respectively.
For comparison, the corresponding differences between LAPW($1s^2$ Dirac PBE/aZORA HF) and LAPW($1s^2$ NR HF/NR HF) in the main text are $-1.11$ and $-1.19$~kcal/mol, respectively.
The two estimates thus agree to within 0.06~kcal/mol, supporting the reliability of the relativistic treatment and quantifying the spread between the tested approximations.

Both, the deviations from using the standard LAPW(PBE/PBE) setup with a reduced cutoff and the difference in relativistic corrections between the tested approximations, are much smaller than the spread of formation energies reported in the literature (Tab.~\ref{tab:si-interstitials}).

\begin{table}[H]
\caption{\label{tab:si-interstitials} Self-interstitial formation energies (in kcal/mol) of Si in the hexagonal (H) and split (X) configurations, obtained in this work and extracted from the literature.
$\Delta E_{\mathrm{H}-\mathrm{X}}$ is a difference between the formation energies of the two interstitial types.
Values in italics are calculated using a $4\times4\times4$
$\mathbf{k}$-point mesh. Values in bold are the final results of this work and represent estimates of the energies on the
$5\times5\times5$ $\mathbf{k}$-point mesh.
}
\centering
\begin{tabular}{llll}
\hline
\hline
Core/basis & $E^{\mathrm{form}}_{\mathrm{H}}$ & $E^{\mathrm{form}}_{\mathrm{X}}$ & $\Delta E_{\mathrm{H}-\mathrm{X}}$\\
\cline{1-4}\\[-8pt]
\multicolumn{4}{c}{\texttt{exciting} (this work)}\\
LAPW (NR PBE/NR PBE)         & \textit{190.43} & \textit{185.11} & \textit{5.32}\\
LAPW (NR HF/NR HF)         & \textit{190.36} & \textit{185.00} & \textit{5.35}\\
LAPW (NR HF/NR HF) + $\mathbf{k}$-corr. & \textbf{190.25} & \textbf{184.96} & \textbf{5.29} \\
LAPW (Dirac PBE/aZORA HF) & \textit{189.24} & \textit{183.81} & \textit{5.43}\\
LAPW (Dirac PBE/aZORA HF) + $\mathbf{k}$-corr. & \textbf{189.14} & \textbf{183.77} & \textbf{5.36} \\
\cline{1-4}\\[-8pt]
\cline{1-4}\\[-8pt]
\multicolumn{4}{c}{Literature}\\
Periodic POB AE/POB-DZVP\textsuperscript{\emph{a}}       & 204.5 & 197.4 & 7.1 \\
Periodic POB AE/POB-TZVP\textsuperscript{\emph{a}} & 196.7 & 177.6 & 19.1 \\
Periodic POB AE/POB-TZVP-REV2\textsuperscript{\emph{a}}  & 196.2 & 185.4 & 10.8\\
Dunning AE/cc-pVDZ\textsuperscript{\emph{a}}            & 199.0 & 191.4 & 7.6 \\
Dunning AE/cc-pVTZ\textsuperscript{\emph{a}} & 194.6 & 188.6 & 6.0\\
Ahlrichs AE/def2-SVP\textsuperscript{\emph{a}}          & 200.6 & 191.6 & 9.0  \\
Ahlrichs AE/def2-TZVP\textsuperscript{\emph{a}}         & 191.4 & 185.2 & 6.2 \\
Ahlrichs AE/def2-TZVPP\textsuperscript{\emph{a}}        & 191.4 & 185.2 & 6.2 \\
Ahlrichs AE/def2-TZVPP\textsuperscript{\emph{a}}  & 191.4 & 185.2 & 6.2\\
ccECP ECP/cc-pVDZ\textsuperscript{\emph{a}} & 204.3 & 195.1 & 9.2\\
ccECP ECP/cc-pVTZ\textsuperscript{\emph{a}} & 191.6 & 185.6 & 6.0\\
ccECP ECP/PW (QE)\textsuperscript{\emph{a}} & 190.5 & 187.0 & 3.5\\
PAW/PW (VASP)\textsuperscript{\emph{b}} & 188.2 & 182.9 & 5.3\\
\hline
\hline
\end{tabular}

\textsuperscript{\emph{a}}~Ref.~\cite{Simula-Si-defects-2026};\\
\textsuperscript{\emph{b}}~Ref.~\cite{Gruneis-2023-Si-defects}.
\end{table}
